\documentstyle[12pt,epsfig]{article}
\newcommand{\be}{\begin{equation}}
\newcommand{\ee}{\end{equation}}
\newcommand{\bea}{\begin{array}{c}}
\newcommand{\eaa}{\end{array}}
\newcommand{\ba}{\begin{eqnarray}}
\newcommand{\ea}{\end{eqnarray}}

\begin{document}
%
%
\date{\today}
\title{Charmonium Production from the Secondary  Collisions at
    LHC Energy }
\author{Peter Braun-Munzinger$^1$ and Krzysztof Redlich$^2$}
\maketitle
\vskip 0.3 true cm
{\centerline {$^1$Gesellschaft f\"ur
        Schwerionenforschung, GSI}}

{\centerline {
        Postfach 110552, D-64220 Darmstadt, Germany }}

{\centerline {
$^2$Institute for Theoretical Physics, University of Wroc\l aw,}}

{\centerline {
PL-50204  Wroc\l aw, Poland}}
\vskip 0.5 true cm
\begin{abstract}
We consider the charmonium production in thermalized hadronic
medium created in ultrarelativistic heavy ion collisions at LHC
energy.
 The calculations for the secondary $J/\psi$ and $\psi^,$ production
  by $D\bar D$ annihilation
 are
 performed within a kinetic model taking into account the space-time
 evolution of a longitudinally and transversely expanding medium.
 We show that the secondary charmonium production appears almost
 entirely during the mixed phase and it is very sensitive  to the
 charmonium      dissociation cross section with co-moving hadrons.
 Within the most likely scenario for the dissociation cross section
of the $J/\psi$ mesons their regeneration in the hadronic medium
 will be negligible. The secondary production of $\psi^,$ mesons
 however, due to their large cross section above the threshold, can
 substantially exceed the primary yield.
\end{abstract}
\section{Secondary charmonium production}

The initial energy density in ultrarelativistic heavy ion collisions
at LHC energy exceeds by a few order of magnitudes the critical value
required for quark-gluon plasma formation. Thus, according to Matsui
and Satz      [1,2], one expects the formation of charmonium bound
states to be severely suppressed due to Debye screening. The initially
produced $c\bar c$ pairs in hard parton scattering, however, due to
charm conservation, will survive in the deconfined medium until the
system reaches the critical temperature where the charm quarks
hadronize, forming predominately $D$ and $\bar D$ mesons.  An
appreciable fraction of $c\bar c$ pairs and consequently $D$,$\bar D$
mesons produced in Pb-Pb collisions at LHC energy can lead to
an additional production of charmonium bound states due to  reactions
such as: $D\bar {D^*} +D^*\bar D + D^*\bar {D^*}\to \psi +\pi$ and
$D^*\bar {D^*} +D\bar D \to \psi +\rho$ as first indicated in
\cite{3}.  In this work we present a quantitative description of
the secondary $J/\psi$ and $\psi^,$ production due to the above processes
from the thermal hadronic medium created in Pb-Pb collisions at LHC
energy.

\section{  Thermal production kinetics}
 The charmonium production cross section
 $\sigma_{D\bar D\to \psi h}$
 can be related to the  hadronic absorption of charmonium
 $\sigma_{\psi h\to D\bar D}$,
 through the detailed balance
 relation
\begin{equation}
\sigma_{D\bar D\to \psi h} = d_{D\bar D} ({k_{\psi\pi }\over {k_{D\bar D}}})^2
\sigma_{\psi h\to D\bar D},
\end{equation}
\begin{figure}[htb]
\begin{center}
\epsfig{file=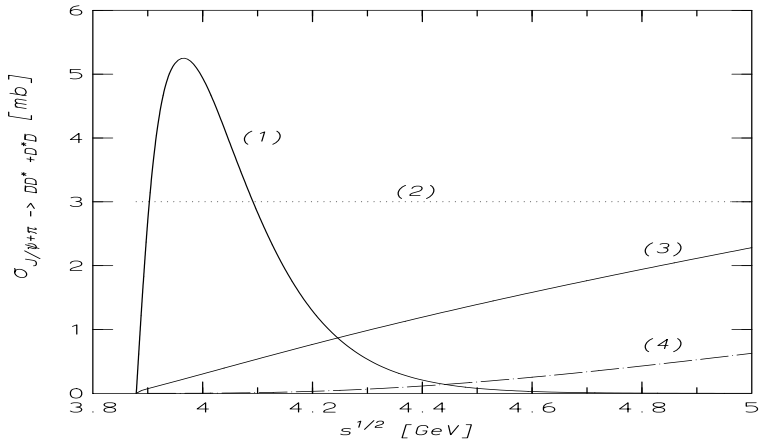, width=110mm,height=64mm}
\end{center}
\vspace*{-0.3 cm}
\caption
      {
 The cross section for $J/\psi +\pi  \to {\overline D}D^*
 +    {\overline {D^*}}D$ absorption
 as a function of energy:  curve (1) is calculated within
  a non-perturbative
 quark exchange model [4];  curve (2) corresponds to  the
  constant cross section of 3 mb [5,6];
  curve (3) was obtained within the framework
  of a meson exchange model [7];
  curve (4) is derived in the context of perturbative
  QCD [8].
 }
\label{fig:1}
\end{figure}
\begin{figure}[htb]
\begin{center}
\epsfig{file=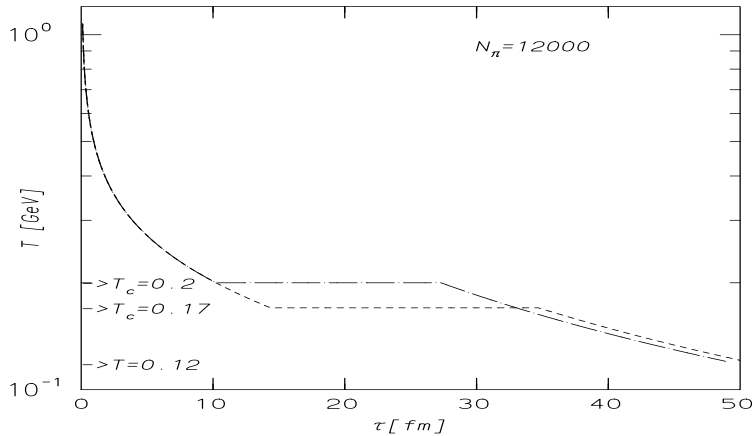, width=110mm,height=64mm}
\end{center}
\vspace*{-0.3 cm}
\caption
      {
Model for time evolution of temperature in the expanding QCD medium
crated in Pb-Pb collisions at LHC energy.
 }
\label{fig:2}
\end{figure}
\begin{figure}[htb]
\begin{center}
\epsfig{file=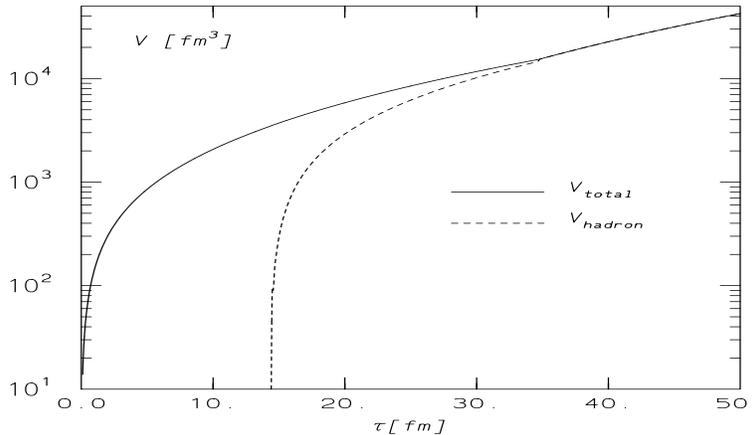, width=110mm,height=64mm}
\end{center}
\vspace*{-0.3 cm}
\caption
      {{\rm
As in fig.1. but for the volume of the system: $V_{hadron}$ is
the fraction of the volume occupied by the confined-hadronic medium;
$V_{total}$ is the total volume of the  system.
 }}
\label{fig:3}
\end{figure}
\begin{figure}[htb]
\begin{center}
\epsfig{file=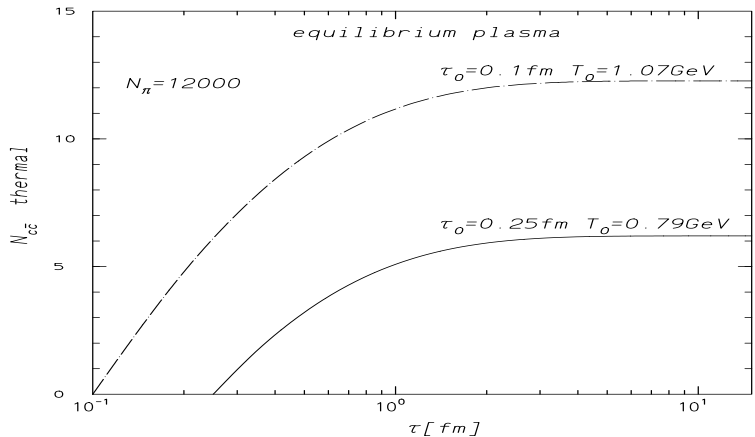, width=110mm,height=64mm}
\end{center}
\vspace*{-0.3 cm}
\caption
      {{\rm
Time evolution of the number of $c\bar c$ produced in an
      equilibrium QGP for two different values of the initial
thermalization time and the initial temperature  calculated
with 12000 pions in the final state.
 }}
\label{fig:4}
\end{figure}
\begin{figure}[htb]
\begin{center}
\epsfig{file=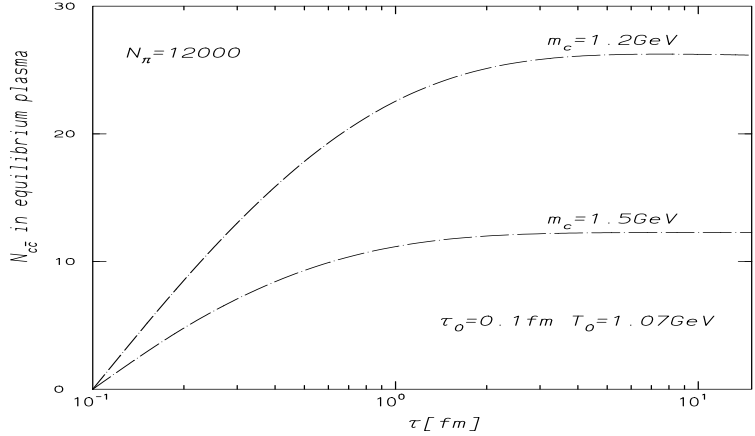, width=110mm,height=64mm}
\end{center}
\vspace*{-0.3 cm}
\caption
      {{\rm
Time evolution of the number of $c\bar c$ produced in an
      equilibrium QGP for two different values of the
charm quark mass.
 }}
\label{fig:5}
\end{figure}
\begin{figure}[htb]
\begin{center}
\epsfig{file=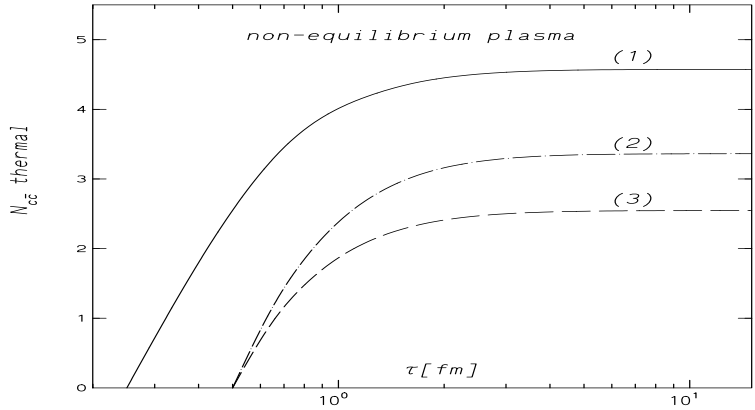, width=110mm,height=64mm}
\end{center}
\vspace*{-0.3 cm}
\caption
      {{\rm
Time evolution of the number of $c\bar c$ produced in out of chemical
      equilibrium QGP calculated with the initial conditions
from SSPC and HIJING.
 }}
\label{fig:6}
\end{figure}
\begin{figure}[htb]
\begin{center}
\epsfig{file=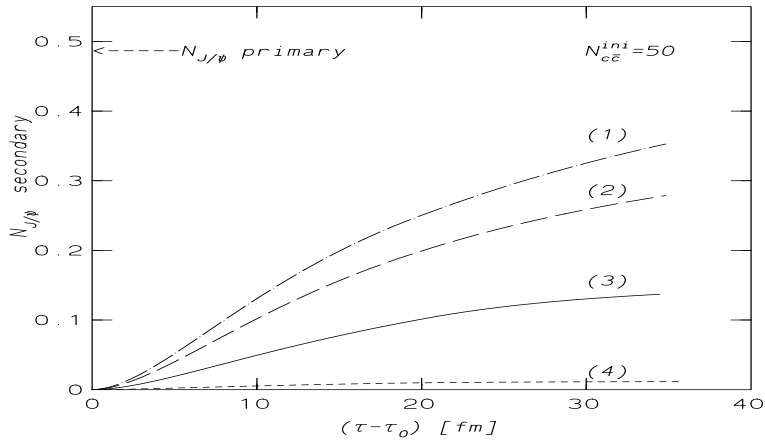, width=110mm,height=64mm}
\end{center}
\vspace*{-0.3 cm}
\caption
       {{\rm
Time evolution of $J/\psi$ production from hadron gas   due to
${\overline D}D^* +D{\overline {D^*}}\to J/\psi \pi$ process
calculated for    Pb-Pb collisions at LHC
energy for four different parameterizations of the $J/\psi - \pi$
absorption cross sections as described in fig. 1.
 }}
\label{fig:7}
\end{figure}
\begin{figure}[htb]
\begin{center}
\epsfig{file=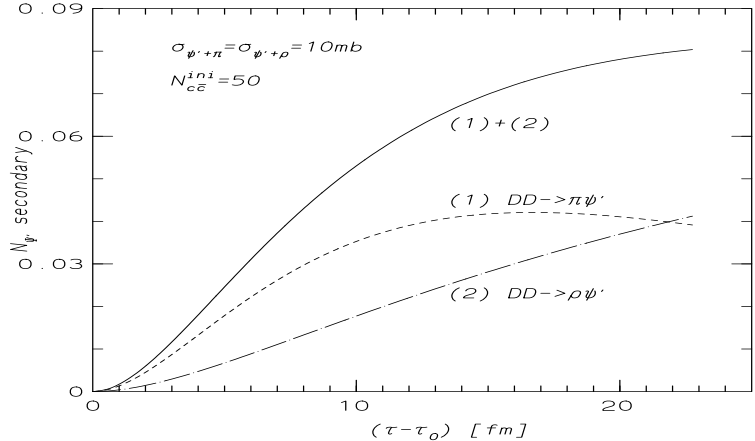, width=110mm,height=64mm}
\end{center}
\vspace*{-0.3 cm}
\caption
{Time evolution of the abundance of $\psi^,$ from Pb-Pb
collisions at LHC using constant absorption cross sections for
$\psi^, - \pi$ and
$\psi^, - \rho$ of 10 mb.
 }
\label{fig:8}
\end{figure}
\begin{figure}[htb]
\begin{center}
\epsfig{file=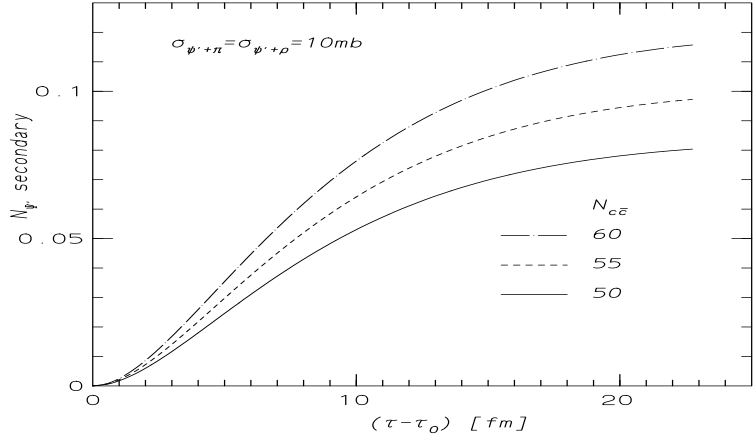, width=110mm,height=64mm}
\end{center}
\vspace*{-0.3 cm}
\caption
{As in fig. 8 but  for different values of the number of $D\bar D$
mesons.
}
\label{fig:9}
\end{figure}
\begin{figure}[htb]
\begin{center}
\epsfig{file=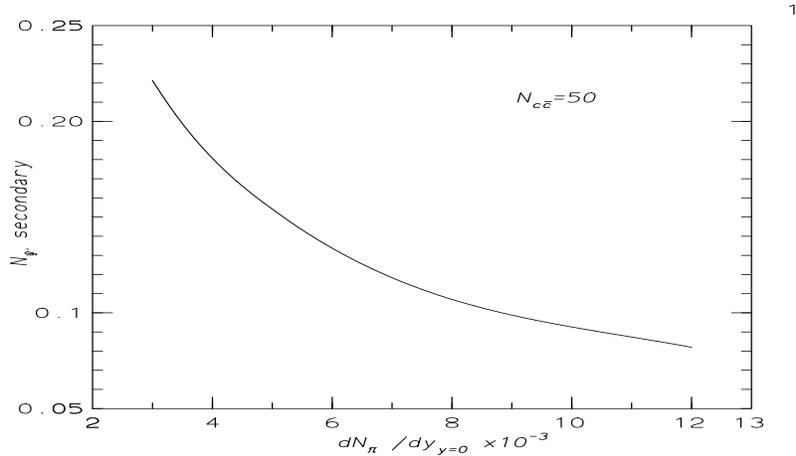, width=110mm,height=64mm}
\end{center}
\vspace*{-0.3 cm}
\caption
{Total $\psi^,$ yield produced during the mixed phase as a function
of  pion multiplicity.
}
\label{fig:10}
\end{figure}
\begin{figure}[htb]
\begin{center}
\epsfig{file=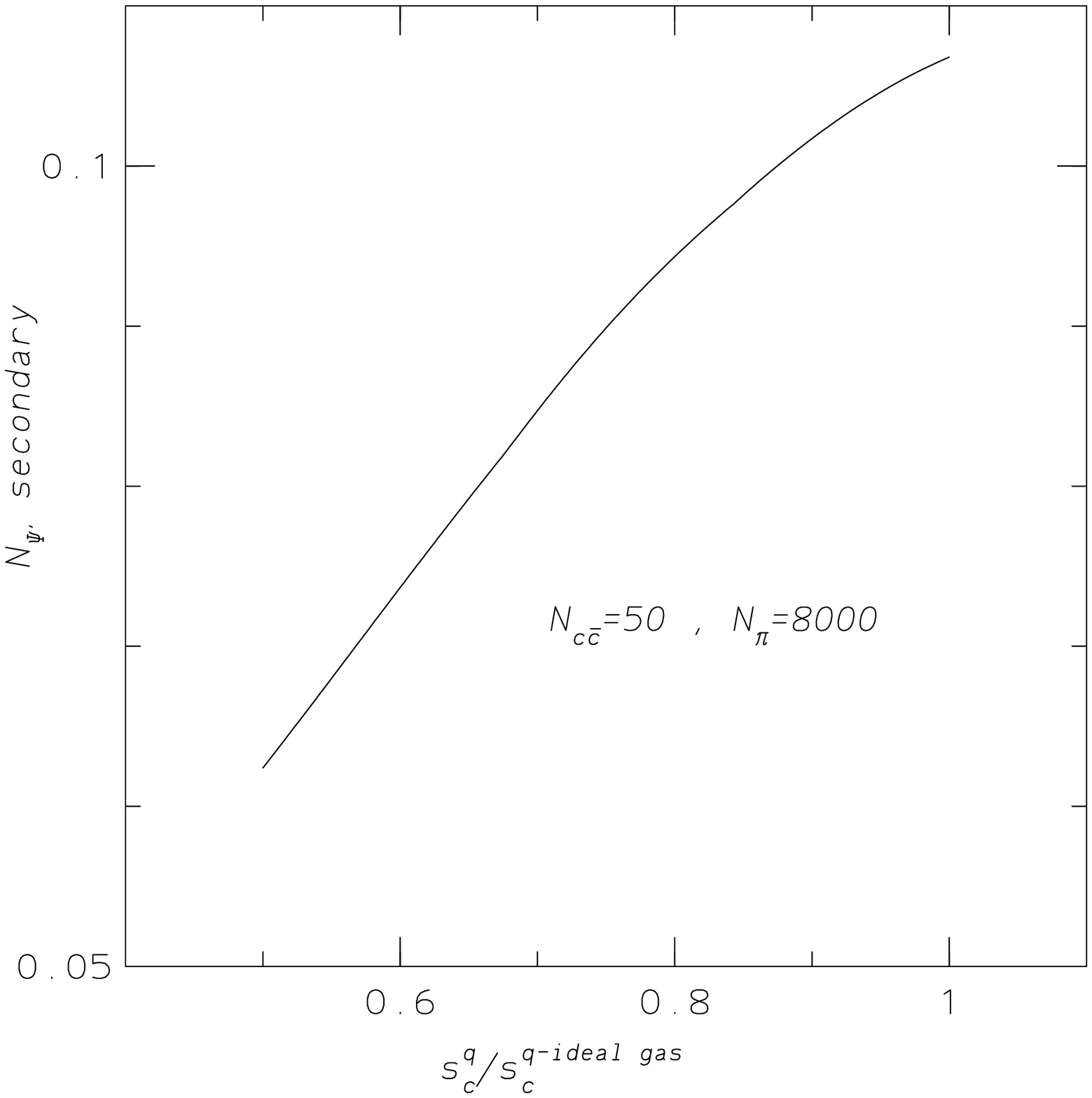, width=110mm,height=64mm}
\end{center}
\vspace*{-0.3 cm}
\caption
{Total $\psi^,$ yield produced during the mixed phase as a  function
of  the fractions of the critical entropy density of
the non-interacting  quark-gluon plasma.}
\label{fig:11}
\end{figure}
\par\noindent
where $k_{ab}^2=[s-(m_a+m_b)^2][s-(m_a-m_b)^2]/4s$ denotes the
square of the center of mass momentum of the corresponding reaction and
$h$ stands for $\rho$ or $\pi$,  whereas $\psi$ for $J/\psi$ or
$\psi^{,}$ mesons. The degeneracy factor $d$ takes the values:
$
d_{D\bar {D^*}}=
d_{D^*\bar D}=3/4$,
$
d_{D^*\bar {D^*}}=
d_{D^*\bar {D}}=1/4$ with the pions in the final state and
$
d_{D\bar {D^*}}=
d_{D^*\bar {D}}=9/4$,
$
d_{D\bar {D}}=27/4,
$
$
d_{D^*\bar {D^*}}=3/4,
$
for the processes with     $\rho$ mesons.

The magnitude of charmonium absorption cross section
 on  hadrons is still, however, theoretically not well
 under control.
 There are  four models which have been proposed
 to give an estimate for the $J/\psi$ dissociation
cross section on pions:
(1) the constituent quark model  \cite{4}, (2) the comover model \cite{5,6},
 (3) an effective hadronic Lagrangian \cite{7}
and  (4) a short distance QCD approach \cite{8}.

In the framework of the constituent quark model the $J/\psi$
absorption on a pion is viewed as a quark exchange process where the
charm quark changes its side with a light quark. The cross section
for the
$\pi +J/\psi\to D^*\bar D + D\bar {D^*} + D^*\bar {D^*}$
process,
in this approach, was calculated  in the
non-relativistic quark model in terms of the first Born approximation \cite{4}.
 The energy dependence of
the above cross section  for the final  state
$D^*\bar D + D\bar {D^*}$  is shown in fig. 1.
One sees that the cross section abruptly increases just above  threshold
and reaches its peak value of about 6 mb.  The large value of
the cross section is mostly due to the particular modelling
of the confining interaction between the quarks which is  taken
as attractive, independent of the colour quantum number of the affected
quark pair.

 The large charm quark mass, $m_c\sim 1.2-1.8$ GeV and $J/\psi$
binding energy, $2M_D -M_{J/\psi} \sim$640 GeV in comparison
to   $\Lambda_{QCD}$, as well as small size of
$J/\psi$, $r_{J/\psi}\sim 0.2$fm$<<\Lambda_{QCD}^{-1}$, have been
used as an argument  to calculate the
charmonium dissociation on light hadrons in terms of a QCD approach
\cite{8,9}.
 The
$J/\psi - h$ cross section in this approach is expressed in terms of
hadronic gluon distribution functions, making use of the short
distance QCD method based on sum rules derived from    the
operator product expansion. The resulting energy dependence
of the $J/\psi$ absorption cross section on pions calculated within
QCD
\cite{8} is shown
in fig. 1. The hadronic gluon distribution functions are strongly
suppressed at high gluon momenta. Thus, since the dissociation of $J/\psi$
requires hard gluons, the absorption cross section $J/\psi \pi\to D\bar D$
becomes very small just above
    threshold as seen in fig. 1.
Recent analysis of $J/\psi $ photoproduction data          confirms
the relation between the charmonium-hadron interactions and the
hadronic gluon structure \cite{8,10,10b} indicating a small value
of the $J/\psi$ absorption cross section on hadrons.

The  scattering of $J/\psi$ on $\pi$ or $\rho$ meson
can be also described as the exchange of open charm mesons between
the charmonium and the incident particle. Near
the kinematical threshold  this exchange  was  modeled in terms
of an effective meson Lagrangian \cite{7}. In fig. 1 we  show
the energy dependence of the cross section for
 $J/\psi$ dissociation in this approach.
The result for the cross section should be considered here as an upper limit
since in the model the hadronic form-factor was taken to be    unity.

In phenomenological models describing     charmonium production
in pA and light nucleus-nucleus collisions the dissociation cross section
of charmonium on light mesons was
 assumed to be energy independent \cite{5,6}.
Using a    value of  $\sigma_{\psi N}\sim 4.8$ mb, which was shown
to be consistent with the pA data,
and the quark
structure of hadrons one could fix the $\sigma_{\psi (\pi ,\rho)}\sim
2\sigma_{\psi N}/3\sim 3$ mb.

The differences between these various  models for the energy dependence of the
$J/\psi$-$\pi$ cross section are  particularly
large near  threshold.
  The  theoretical uncertainties
 of the cross section seen in fig. 1
 will naturally influence the yield of
 the secondary charmonium production.
 In the following, we shall calculate and compare the yield
 of the secondary
 $J/\psi $ with all  these cross sections.

The $\psi^,$ absorption cross section has not been evaluated
in the quark exchange and effective Lagrangian model.
The application of the short distance QCD is not possible in this case
due to the very low binding energy of the $\psi^,$ ( of the order of
60 MeV) and the correspondingly large size of the $\psi^,$.

 For a first estimate of the $\psi^,$ dissociation on light mesons
we assume that the absorption cross section is energy independent
and attains its geometric value of about 10 mb very near the
threshold. This assumption was shown to  provide a quite good
description of the $\psi^,$ suppression observed in S-U collisions
\cite{11}.

\subsection{Rate equation for the charmonium production in a hadron gas}

In  a   thermal hadronic medium  the rate of charmonium production
 from $D\bar D$ annihilation is determined by
the thermally averaged cross section
and the densities of incoming and outgoing particles.
The thermal average of the charmonium production cross section
$<\sigma_{D\bar D\to \psi h}v_{D\bar D}>$ is given by the following
expression \cite{11b}:
\begin{equation}
<\sigma_{D\bar D}v_{D\bar D}> =
{\beta\over 8}
{
{\int_{t_0}^\infty   dt
 \sigma_{D\bar D}(t)
 [t^2-(m_{D\bar D}^+)^2]
 [t^2-(m_{D\bar D}^-)^2]
 K_1(\beta t)}\over
 {m_D^2m_{\bar D}^2K_2(\beta m_D)K_2(\beta m_{\bar D})}
},
\end{equation}
where  $K_1$, $K_2$ are modified Bessel functions of the second kind,
$m_{D\bar D}^+\equiv m_D+m_{\bar D}$ and
$m_{D\bar D}^-\equiv m_D-m_{\bar D}$,
$t\equiv \sqrt {s}$ is the center-of-mass energy, $\beta$ the
inverse temperature,
$v_{ab}$ is the relative velocity of incoming mesons and the
integration limit is taken to be: $t_0=max
[(m_D+m_{\bar D}),(m_\psi+m_h)]$.

The thermal cross section for the production and absorption
of charmonium  convoluted with the densities of
incoming particles describes  the rate equation for charmonium
production in a hadron gas per unit of rapidity by:

\begin{equation}
{{dR}\over {d\tau }}=
\sum_{i,j}<\sigma_{D_i\bar D_j\to \psi h}v_{D_i\bar {D_j}}>
n_{D_i}n_{\bar {D_j}}
-
\sum_{i,j}<\sigma_{\psi h \to D_i\bar {D_j}}v_{\psi h}>
n_{\psi }n_{h},
\end{equation}
where $i=\{D, D^*\} $,
$h$ denotes a $\pi$ or $\rho$ meson and $\psi$ stands for  $\psi^,$ or $J/\psi$.

In the hadronic
medium we include the following secondary charmonium production processes:
\begin{equation}
D^*\bar D + D\bar {D^*} + D^*\bar {D^*}
\to
\pi +\psi
\end{equation}
\begin{equation}
D^*\bar {D^*} + D\bar {D}
\to
\rho +\psi
\end{equation}
where $\psi$ stands for $J/\psi$ or $\psi^,$.

The solution of the rate equation requires additional assumptions
on    the space-time evolution of the hadronic medium  and the initial
number of $D$ and  $\bar D$ mesons.

\subsection{ Model for expansion dynamics}

We have adopted a
 model for the expansion dynamics
assuming isentropic cylindrical expansion of the thermal fireball.
 To account for transverse expansion we assume  a
 linear increase of the  radius of a   cylinder with  the proper time.
Thus, at a given proper time
 $\tau $ the volume of the system is parameterized by:
\begin{equation}
 V(\tau )=\pi R^2(\tau ) \tau
 ~~{\rm with}~~
  R(\tau )=R_A + 0.15 (\tau -\tau_0)
\end{equation}
 where
  $\tau_0$ is the initial
 proper time when  the system is created as a thermally equilibrated
 quark-gluon plasma and   $R_A$ is an initial radius being determined
 by the atomic number of colliding nucleus. For central Pb-Pb collisions
 $R_A\sim 6.7$ and we fixed  $\tau_0\sim 0.1$ fm.

 The initially produced quark-gluon plasma
cools during the expansion until it reaches
 the critical temperature $T_c$ at the time
$\tau_q^c$ where it starts to hadronize.
 The system  stays in the mixed phase until the time $\tau_h^c$, where
 the quark gluon plasma is totally converted to a hadron gas.
 The  purely hadronic gas  can still expand until
 it reaches the chemical freeze-out temperature $T_f$ at the time $\tau_f$
 where all inelastic particle scattering ceases and  the particle
 production    stops.

 The most recent results of Lattice Gauge
 Theory  (LGT) give an  upper limit for the critical temperature
 in QCD to be $T_c\sim 0.17$GeV
 \cite{12}. The analysis of presently
 available data for different particle
multiplicities and their ratios measured
 in heavy ion collisions at SPS energy suggests that the chemical
 freezeout temperature
at LHC  should be in the range
 $0.16<T_f<0.18$ \cite{13,13a,14,15}, that is very close to $T_c$.

To make a quantitative description of the space-time evolution
 of a thermal medium one still needs to specify  the
equation of state.
 The quark gluon plasma
is considered as  an ideal gas of  quarks and gluons whereas
the hadron gas is described as an ideal gas of hadrons and resonances.
We have included the contributions of all baryonic and mesonic resonances
with a   mass of up to 1.6 GeV into the partition function.
To  take into account
 approximately
the repulsive interactions between hadrons at short distances
we apply    excluded volume corrections.  Here we use the thermodynamically
consistent model proposed in \cite{14,16} where the thermodynamical
observables for extended particles are obtained from the
formulas for pointlike objects but with the shift of the
chemical potential. In particular the pressure is described by:
\begin{equation}
 P^{extended}(T,\mu )=P^{pointlike}(T,\bar\mu),
~~~ {\rm with}~~~\bar\mu=\mu -v_{eigen}p^{extended}(T,\mu),
\end{equation}
where the particle eigenvolume $v_{eigen}=4{4\over 3}\pi r^3$
simulates repulsive interactions between hadrons. Following the detailed
analysis in \cite{14,17} we have assigned the value $r=0.3$ fm  for all
mesons and baryons.

In fig.2 we show the time evolution of the temperature
for Pb-Pb collisions at LHC energy for two different values of $T_c$.
The initial temperature of the chemically equilibrated plasma
 was fixed  requiring  the  entropy conservation:
\begin{equation}
s_{q}(T_0)V_0\sim N_{\pi }/3.6 ~~{\rm with}~~
N_\pi\equiv (dN_{\pi }/dy)_{y=0},
\end{equation}
where $N_\pi$ described the final number of pions  at midrapidity,
$V_0$ is the initial volume  and $s_q(T_0)$  the initial entropy density
of the thermalized quark-gluon plasma.
 For
$N_\pi =12000$ and $\tau_0$=0.1 fm we get the initial
 thermalization temperature  $T_0\sim 1.07$GeV for Pb-Pb collisions
 at LHC.

The result in fig.2 shows that for $T_c\sim 0.17$GeV
the QGP starts to hadronize after
14 fm and stays in the mixed phase by 20 fm. Taking the lowest expected
freezeout  temperature
$T_f\sim 0.16$ GeV the system exists only for a very short time of
 few  fm in a hadronic phase.\footnote{we do not
 count the isentropic expansion phase towards thermal freeze-out,
 since no appreciate particle production takes place there.}
 Thus, the contribution of the purely
hadronic phase to the overall secondary charmonium multiplicity
  can be neglected.
The secondary charmonium bound states are produced almost
entirely  during
the mixed phase. This is in contrast to \cite{3}.
In fig. 3 we show the  time evolution of the total
volume $V$ as well as the fraction of the
volume $V_h$ occupied by  hadrons during the mixed phase.
The total volume is calculated following eq.6, whereas $V_h$ can be
obtained from the condition of entropy conservation in the following
form:
\begin{equation}
V_h(\tau )={{V(\tau )-V_q^c}\over {1-{{s_h^c}\over {s_q^c}}}} ~~, ~~
V_q^c={{N_\pi}\over {3.6s_q^c}}~~,~~
V_h^c={{N_\pi}\over {3.6s_h^c}},
\end{equation}
where $s_q^c$ and $s_h^c$ are  the entropy density in the
quark gluon plasma and the hadron gas at the critical temperature.
For the  equation of state described above and   $T_c=0.17$GeV
we have: $s_q^c\sim 11.9$ and $s_h^c/s_q^c\sim 0.22$.

 The       time evolution of the hadronic medium during the mixed phase
                            is
totally determined by the number of pions in the final state
and the entropy density of the quark-gluon plasma and the hadron gas at
the critical temperature.
Our default parameters for the expansion dynamics are as follows:
\begin{equation}
T_c=0.17{\rm GeV} ~,~ T_f\sim T_c ~,
~ \tau_0\sim 0.1 {\rm fm}~,~ s_q^c\sim 12/{\rm fm^3}~,~
        s_h^c/s_q^c\sim 0.22.
\end{equation}
We will study, however, how deviations from the above values
influence  the final multiplicity of the secondary charmonium
produced during the evolution of      hadronic medium.

\subsection{Charm density}

The initial number of open charm mesons
at the critical  temperature  $N_{D\bar D}$
 can be related to
 the number
of primary $c,\bar  c$ quarks
$N_{c\bar c}$
produced in A-A collisions.
Neglecting the possible absorption and
 production of $c\bar c$ pairs during
the evolution and hadronization of a quark-gluon plasma we can put
$N_{D\bar D}\sim
N_{c\bar c}$. Due to  charm conservation
the number of $D^,$s  should be equal to the number
of $\bar D$ mesons. The charm conservation also
implies that during the mixed phase the number  of $D\bar D$
mesons in the hadronic  phase
 $N_{D\bar D}^{m}$is given by the fraction of     volume
occupied by the hadrons that is $N_{D\bar D}^{m}=N_{c\bar c}V_h/V$
where $V$ and $V_h$ are  described by eq.6 and 9.
We further  assume that
during      hadronization of the quark gluon plasma the open
charm mesons are in a local thermal equilibrium with all other
hadrons, however,            the yield of $D$ and $\bar D$ exceeds
their chemical equilibrium value.
The ratio
of the number of $D$ to  $D^*$ mesons at the temperature
 $T$ is obtained from the  relative chemical equilibrium  condition:
\begin{equation}
{{N_D}\over {N_{D^*}}}=3
{{ m_D^2K_2(\beta m_D)}\over
{m_{D^*}^2K_2(\beta m_{D^*})}}.
\end{equation}

 The initial number of $c\bar c$ pairs in  Pb-Pb collisions
at LHC energy was obtained by scaling the p-p result with the geometrically
allowed number of nucleon-nucleon collisions.  The cross section
for charm production at LHC in p-p collisions was calculated by leading
order perturbative QCD
using code PYTHIA with $<k_t^2>=1$GeV$^2$ \cite{18}.
Typical values of the 50$c \bar c$ pairs and $0.5$$J/\psi$ were obtained
in central Pb-Pb collisions at midrapidity.

 The $c\bar c$ pairs can be also produced and absorbed during the
evolution of thermally equilibrated quark-gluon plasma. In  lowest
 order in $\alpha_s$ the charm quark pairs are produced by
   gluon  and quark pair fusion. Solving a similar rate equation
as in eq.3 \cite{19} but with the appropriate QCD cross sections
we calculated
 in fig.4  the time evolution of the number of thermal $c\bar c$ pairs.
The  thermal production is seen in fig.4 not to be
 negligible as compared with the
preequilibrium production. Dependent   on the
thermalization time there are an additional 5-10 $c\bar c$ pairs
produced during the evolution of the thermally and chemically
equilibrated plasma. This number, however, depends crucially on the
value of the charm quark mass $m_c$ and the strong coupling constant  $\alpha_s$.
In fig.5 we compare the thermal $c\bar c$ multiplicity calculated with
$m_c=1.5$ GeV and $m_c=1.2$ GeV with  $\alpha_s=0.3$.

Dynamical models for parton production and evolution like
HIJING \cite{20,21} or SSPC \cite{22} confirmed, that
indeed, after a very short time
of the order of 0.1-0.5fm the partonic medium can reach the  thermal
equilibrium. However, both quarks and gluons my appear far below
their chemical saturation values. Thus, discussing thermal $c\bar c$
production one should take into account deviations of
gluon and light quark yields from their equilibrium values.
These deviations can be  parameterized by    fugacities parameters
$\lambda_q$ and $\lambda_g$ modifying
the Boltzman distribution function of quarks and gluons \cite{23}.

In fig.6 we  calculated the time evolution of $c\bar c$ pairs
within the kinetic model derived in \cite{23,20} for a non-equilibrium
quark-gluon plasma. The following initial conditions for the thermalization time
$\tau_0$, the initial temperature $T_{0}$ and the values for
quark $\lambda_q$ and gluon $\lambda_g$ fugacities
were fixed for Pb-Pb
collisions at LHC from     SSPC \cite{22} and HIJING \cite{21,24}:
(1) $\tau_{0}=0.25fm$, $T_{0}=1.02GeV$, $\lambda_g^{0}=0.43$
$\lambda_q^{0}=0.082$  \cite{22};
(2) $\tau_{in}=0.5fm$, $T_{in}=0.82GeV$, $\lambda_g^{in}=0.496$
$\lambda_q^{in}=0.08$  \cite{21};
(3) $\tau_{in}=0.5fm$, $T_{in}=0.72GeV$, $\lambda_g^{in}=0.761$
$\lambda_q^{in}=0.118$  \cite{24}.
In a non-equilibrium plasma, dependent on the initial conditions,
there are  3-5
 $c\bar c$
pairs produced during the evolution of     plasma as seen in fig.6.
 One should
also note that at the time when     temperature reaches the critical
value of $0.17$GeV the quark and gluon  fugacities are
very close  to     unity which indicates the chemical equilibration of
an ideal quark-gluon plasma.

From the results presented in fig$^,$s.4-6 one     concludes, that
dependent
on the initial conditions as well as on the amount of chemical
equilibration of the  plasma, one   expects 5-10
$D\bar D$ pairs at $T_c$ in addition to the 50 $D\bar D$ from the
hadronization of the initially produced $c\bar c$ pairs.

\section{Time evolution of the charmonium abundances}

The number of produced charmonium bound states  from $D\bar D$
scattering is obtained by multiplying the rate equation eq.3
by the volume of the hadron gas  eq.9 and then performing the time
integration. In fig.7 we show the time evolution of the abundance
of $J/\psi$ from Pb-Pb collisions at LHC energy obtained by solving
the rate equation.
The calculations were done with the four different models for
$J/\psi-\pi$ absorption cross section described in fig.1 and
 assuming
initially 50 $D\bar D$ and 12000 pions at     midrapidity.
The results in fig.7 show a very strong
sensitivity of $J/\psi$ production  on the absorption cross section.
The largest number of the secondary $J/\psi $ is obtained with the
cross section predicted by the quark exchange model. Here
the number of $J/\psi $ is only by 1/2 smaller than  the primary value.
If, however, the absorption cross section is described by short
distance QCD the secondary $J/\psi$
 production is lower by more than two orders of
magnitude.

 Recent analysis of
$J/\psi$ photoproduction data  confirms the relation between the
energy dependences of the $J/\psi $ cross section  and the Feynmann x-dependence
of the gluon distribution function of the nucleon \cite{8,10}. This could be used
as an indication   that the
 short distance QCD approach is  a consistent way to calculate the $J/\psi$
cross section. Thus,  the
 secondary production of $J/\psi$ from the hadronic gas can be
entirely neglected in this case.
The situation  can, however, change  when discussing the
production of $\psi^,$

The time evolution of the abundance of $\psi^,$ is  presented
in fig.8 with  the same basic parameters as used for $J/\psi$
in fig.7.  We      show  in fig.8 the separate contributions of
 the production processes for the secondary
$\psi^,$  with  $\pi$ and $\rho$ mesons in the final state.
Following the arguments of the previous section the absorption
cross section for $\psi^,$ on $\pi$ and $\rho$ mesons were
taken to be energy independent and equal to its geometric value of  10 mb.
The produced $\psi^,$ during the mixed phase from  $D\bar D$
annihilation is seen in fig.8 to saturate at the large value 0.08,
which is almost 1/5 of the primary number of $J/\psi$ expected
for Pb-Pb collisions at LHC energy.

 The yield of the secondary charmonium
summarized in fig$^,$s.7-8 depends on the parameters
used in     calculations. In the following we study the modification
of the  results on the secondary $\psi^,$
by changing  the initial number of $D\bar D$ mesons
as well as the values of the relevant thermal parameters.

In fig.9 we discuss  the yield of   $\psi^,$
for different multiplicities of $D\bar D$ mesons by including
the contribution of thermal $c\bar c$ pairs.
From the rate equation eq.3 it is clear that this dependence
is quadratic. The increase of $D\bar D$ by 10-20$\%$ implies the
$20-40\%$ increase of the $\psi^,$ yield.

In the  space-time evolution model  all parameters were
fixed assuming  $N_\pi=12000$.
This number, however, is still not
well established  and the deviations for
$N_{\pi}$ in the range
$5000<N_{\pi}<12000$ are  not excluded. In fig.10 we show
the sensitivity of the results for $\psi^,$ production on $N_\pi$.
Keeping the same number of $D\bar D$
mesons and decreasing
$N_{\pi}$  the number of the secondary $\psi^,$ increases as seen
in fig.10. This is mostly because by decreasing
$N_{\pi}$ the  initial
density $n_{D\bar D}$ of $D\bar D$ increases leading to
a larger production of the charmonium bound states.

Modelling the equation of state we have fixed  the  critical
entropy density $s_q^{c-ideal}$ in the quark gluon plasma
by the ideal gas equation of state. From the  Lattice Gauge
Theory we know, however that, due to interactions, the
 entropy density $s_q^c$ could  be  smaller by a significant
factor. Deviation of the momentum distribution of quarks
and gluons from the chemical saturation is also reducing the
critical entropy density of quark-gluon plasma.
To establish the influence of a
 decreasing critical entropy
density we calculated the total number of $\psi^,$ as a function
of the ratio $s_q^c/s_q^{c-ideal}$ in fig.11. Decreasing the
entropy density of a quark-gluon plasma by a factor of two reduces the
yield of $J/\psi$ by 70$\%$. This
is mostly because the time the system
 spends in the mixed phase is shorter. However,
 even with this  reduction of the entropy density
the secondary production of $\psi^,$ is not negligible.
The influence of the result on the  critical entropy density
in a hadron gas $s_h^c$ is less important. Increasing $s_h^c$,
by a factor of two reduces the total number of $\psi^,$ by only 10$\%$.

The secondary $\psi^,$ production in a mixed phase is also sensitive
to the parameterization of the time evolution of the system size
in transverse direction.
Increasing  the  transverse expansion parameter from
 0.15 to $0.45$ in eq.6  decreases  the total number of $\psi^,$ produced
 during the mixed phase
by 40 $\%$. Taking a   quadratic dependence of $R(\tau )$ on
 proper time as proposed
in \cite{3} decreases the yield of $\psi^,$ from 0.08 to 0.036,
 which is still comparable
 with the  number of primary $\psi^,$.

%
%
\section{Conclusions}
We have considered the possibility of the secondary charmonium
production in ultrarelativistic heavy ion collisions
at LHC energy. Admitting thermalization of a partonic medium
crated in a collision and the subsequent  first order phase transition
to a hadronic matter we have
shown that the secondary charmonium production
 appears almost entirely during the mixed phase.  The yield
of secondarily produced $\psi$ mesons is very sensitive to the
hadronic absorption cross section. Within the context of the short
distance QCD approach this leads to negligible values for J/$\psi$
regeneration.  The $\psi^,$ production, however, can be large and may
even exceed the initial yield from primary hard scattering.
Thus it is conceivable that at LHC energy the $\psi^,$ charmonium state can
be seen in the final state whereas $J/\psi$ production can be entirely
suppressed. The appearance of the $\psi^,$ in the final state
could be thus considered as an   indication  for the charmonium
production from the secondary hadronic rescattering.

\vskip 0.2 true cm
\par\noindent
{\bf Acknowledgments}

We  acknowledge stimulating discussions with H. Satz
and J. Stachel. One of us (K.R.)  acknowledges partial support
of the Gesellschaft f\"ur Schwerionenforschung (GSI) and the
Committee of Research Development (KBN).

\end{document}